\documentclass{naturepp}
\usepackage{graphicx}

\title{The Solar Model Problem Solved by 
the Abundance of Neon in Stars of the Local Cosmos} 

\author{Jeremy J. Drake$^{1}$ \& Paola Testa$^2$}

\begin{document}

\maketitle

\begin{affiliations}
 \item Harvard-Smithsonian Center for Astrophysics, 60 Garden
 Street, Cambridge MA 02138, USA.
 \item MIT Kavli Institute for Astrophysics and Space Research, 
Massachusetts Institute for Technology, 70 
Vassar Street, Cambridge, MA 02139, USA.
\end{affiliations}

\begin{abstract}

The interior structure of the Sun can be studied with great accuracy
using observations of its oscillations, similar to seismology of the
Earth.  Precise agreement between helioseismological measurements
and predictions of theoretical solar models\cite{Bahcall.etal:05} has
been a triumph of modern astrophysics.  However, a recent downward
revision by 25-35\%\ of the solar abundances of light elements such as
C, N, O and Ne\cite{Asplund.etal:04} has broken this accordance:
models adopting the new abundances incorrectly predict the depth of
the convection zone, the depth profiles of sound speed and density,
and the helium abundance.\cite{Basu.Antia:04,Bahcall.etal:05} The
discrepancies are far beyond the uncertainties in either the data or
the model predictions.\cite{Bahcall.etal:05b} Here we report on neon
abundances relative to oxygen measured in a sample of nearby 
solar-like stars
from their X-ray spectra.  They are all very similar and substantially
larger than the recently revised solar value.  The neon abundance in
the Sun is quite poorly determined.  If the Ne/O abundance in these
stars is adopted for the Sun the models are brought back into
agreement with helioseismology
measurements.\cite{Antia.Basu:05,Bahcall.etal:05c}

\end{abstract}

The role of the Sun as a fundamental benchmark of stellar evolution
theory, which itself underpins much of astrophysics, renders the
``solar model problem'' one of some importance.  The schism between
helioseismology and models with a revised composition has arisen
because abundant elements such as C, N, O and Ne provide major
contributions to the opacity of the solar interior, which in turn
influences internal structure and the depth at which the interior
becomes convective.  Uncertainties in the calculated opacities
themselves appear insufficient to bridge the
gap,\cite{Bahcall.etal:04,Antia.Basu:05} and the propriety of the
recent abundance revisions has therefore been questioned.

The revised abundances are demanded by new analyses of the visible
solar spectrum that take into account convection and associated
velocity fields and temperature inhomogeneities through 3-D
hydrodynamic modelling, and that relax the assumption of local
thermodynamic equilibrium for computing atomic level
populations.\cite{Asplund.etal:00} The solar abundances of C, N and O
can be measured accurately based on their absorption lines.  However,
Ne lacks detectable photospheric lines in cool stars like the Sun, and
the Ne abundance is therefore much less certain.  The solar Ne content
is assessed based on observations of neon ions in nebular and hot star
spectra, and on measurements of solar energetic
particles.\cite{Meyer:85,Asplund.etal:04} Measurements are generally
made relative to a reference element such as O; the downward revision
of the solar O abundance therefore required a commensurate lowering of
the Ne abundance for consistency.\cite{Asplund.etal:04} Expressed as
the ratio of the number of Ne atoms in the gas to those of O, this
abundance is $A_{Ne}/A_{O}=0.15$, which is very similar to values
adopted in earlier
studies.\cite{Anders.Grevesse:89,Grevesse.Sauval:98} However, it has
recently been pointed out that the solar model problem could be solved
were the true solar Ne abundance to be at least a factor of 2.5 times
higher than recently assessed.\cite{Antia.Basu:05,Bahcall.etal:05b}

\begin{figure}
\center\includegraphics[width=1.0\textwidth]{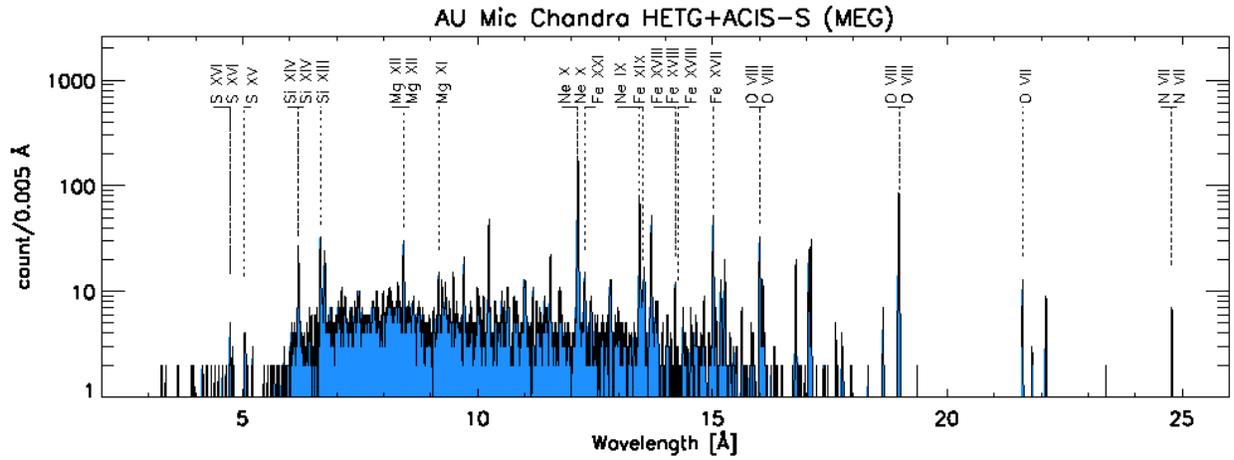}
\caption{A {\it Chandra} Medium Energy Grating X-ray spectrum of the M1~V
star AU Mic.  These data are typical of the {\it Chandra} MEG
spectra upon which this study is based.  Emission line features
superimposed on the bremsstrahlung continuum are
formed predominantly by radiative decay of transitions excited by
electron impact.  Prominent lines are labelled, including the lines of
O~VIII $2p\, ^2P_{3/2,1/2}
\rightarrow 1s\, ^2S_{1/2}$ (18.97~\AA ), Ne~IX $1s2p\, ^1P_1
\rightarrow 1s^2\, ^1S_0$ (13.45 \AA ) and Ne~X $2p\, ^2P_{3/2,1/2}
\rightarrow 1s\, ^2S_{1/2}$ (12.13 \AA ) used here to determine the
Ne/O abundance ratio.}
\label{f:megspec}
\end{figure}

\begin{figure}
\center\includegraphics[width=1.0\textwidth]{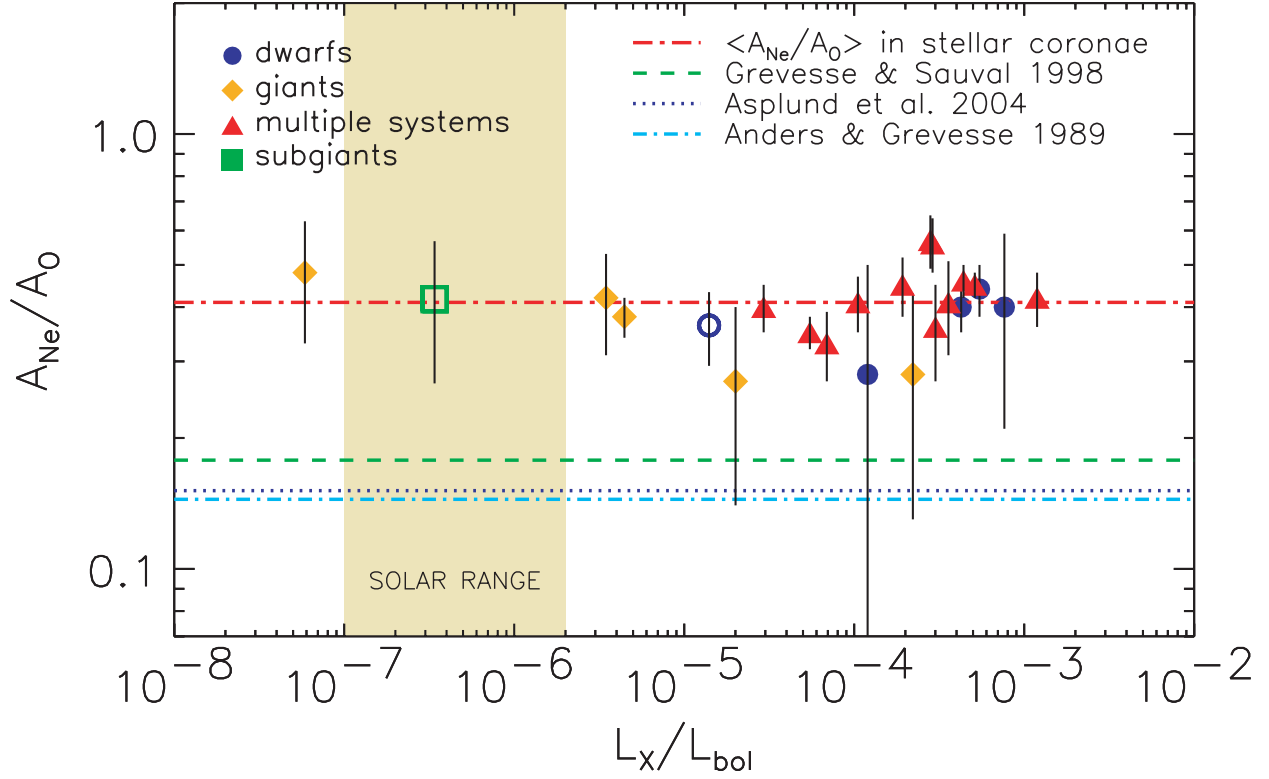}
\caption{Derived Ne/O abundance ratios by number, $A_{Ne}/A_{O}$,
vs.\ the coronal activity index $L_X/L_{bol}$.  Error bars represent
quadrature addition of $1\sigma$ uncertainties of line flux measurement.  Also
shown using hollow symbols are literature
values\cite{Sanz-Forcada.etal:04} for the stars Procyon (F5~IV) and
$\epsilon$~Eri (K2~V) observed using the {\it Chandra} Low Energy
Transmission Grating Spectrometer (LETGS) to better represent the
lower ranges of coronal activity.  The error-weighted mean Ne/O
abundance ratio is $A_{Ne}/A_{O}=0.41$, or 2.7 times the currently
assessed value\cite{Asplund.etal:04} which is illustrated by the
dashed horizontal line.  The recommended value from comprehensive
earlier assessments in common
usage\cite{Anders.Grevesse:89,Grevesse.Sauval:98} are also
illustrated.}
\label{f:abunrats}
\end{figure}

We are motivated by the solar model problem to investigate Ne
abundances in nearby stars.  While not detected in the optical spectra
of cool stars, emission lines of highly ionised Ne are prominent
features of their X-ray spectra.\cite{Drake.etal:01} The Ne/O
abundance ratio can be derived directly from the ratio of observed
fluxes of the hydrogen-like and helium-like ions of O and Ne (see
Methods for details); we adopt a slightly refined version of the
method applied to the analysis of earlier solar X-ray
spectra.\cite{Acton.etal:75} In this way, we have obtained the Ne/O
abundance ratios for a sample of 21 stars lying within 100pc of the
Sun that have been observed by the {\it
Chandra} X-ray Observatory using the High Energy Transmission Grating
Spectrometer.\cite{Canizares.etal:00}
A representative X-ray spectrum,
that of the M1~V star AU~Mic, is presented in Figure~\ref{f:megspec}.
The stars studied, together with observed Ne and O line fluxes and
derived Ne/O abundance ratios are listed in Table~1.  The Ne/O
abundance ratios are illustrated as a function of the ratio of
logarithmic X-ray and bolometric luminosities, $L_X/L_{bol}$---a
commonly used index of stellar coronal activity---in
Figure~\ref{f:abunrats}.  Since our star sample contains more objects
toward higher $L_X/L_{bol}$, we have added Ne/O ratios for two stars
of somewhat lower activity level (Procyon, an F5 subgiant, and
$\epsilon$~Eri, a K2 dwarf) from the
literature.\cite{Sanz-Forcada.etal:04} There is no trend in the Ne/O
abundance ratio with $L_X/L_{bol}$, and the error-weighted mean ratio
is $A_{Ne}/A_O=0.41$.

Our derived Ne/O abundance ratio is 2.7 times higher than the
currently recommended solar value\cite{Asplund.etal:04} but is
consistent with the abundance inferred from
helioseismology.\cite{Antia.Basu:05,Bahcall.etal:05c} Solving the
solar model problem by raising the Ne abundance alone would require a
minimum ratio $A_{Ne}/A_O=0.52$, or 3.44 times larger than
recommended.\cite{Antia.Basu:05} However, raising the C, N, O and Fe
abundances upward within their estimated uncertainty range of $\pm
12$\%\ (adjusting them all together is not unreasonable because the
recent downward revisions are correlated) would require an Ne
abundance higher by only a factor of
2.5\cite{Antia.Basu:05,Bahcall.etal:05c}---quite within our estimated
range.

Extensive review articles on the coronae of the Sun and stars document
evidence that coronal chemical compositions 
often appear different to those thought to characterise the underlying
photospheres.\cite{Meyer:85,Feldman.Laming:00,Drake:03b,Gudel:04} The
differences appear to relate to element first ionisation potentials
(FIPs): the solar corona appears enhanced in elements with low FIP
($\leq 10$~eV; e.g., Mg, Si Fe),\cite{Meyer:85,Feldman.Laming:00} 
whereas much more coronally active
stars appear to have depletions in low FIP elements and perhaps
enrichments of high FIP elements ($\leq 10$~eV; e.g., O, Ne,
Ar).\cite{Feldman.Laming:00,Drake:03b,Gudel:04} At 
first sight, this chemical fractionation might render direct
interpretation of coronal abundance ratios in terms of the composition
of the underlying star problematic.  However, that we are seeing the
same Ne/O ratio in a wide variety of stars sampling a large range of
different coronal activity levels indicates that there is no
significant fractionation between O and Ne in disk-integrated light
from stellar coronae.  This leads us to conclude that the results
represent the true Ne/O abundance ratios of these stars.  This
conclusion is bolstered by findings of a constant Ne/O abundance
ratio, in good agreement with our value, for a small sample of single
and active binary stars observed by {\it
XMM-Newton}.\cite{Audard.etal:03,Gudel:04} In view of the consistency of the
Ne/O abundance ratio in nearby stars, it seems likely that the solar
ratio should be similar.  

There are no recent full-disk integrated light Ne/O measurements for
the Sun; existing studies are instead based on observations of
particular regions and structures of the solar outer atmosphere.  We
provide a tabular summary and bibliography of some of the different Ne/O
estimates since 1974 as Supplementary Information to this letter.  While the
preponderance of estimates
appear consistent with current
assessments, the underlying
observations do not sample photospheric material.  The Ne/O ratio is in
fact observed to differ substantially between the different observations.  In
particular, the highest measured Ne/O ratios based on X-ray lines 
are 2-3 times the accepted solar value and are compatible with the
abundance ratio we find for nearby stars.  These higher Ne/O ratios
were obtained for hotter
active regions\cite{Acton.etal:75,Schmelz.etal:96} 
that are likely to dominate the solar full-disk X-ray
emission.  These measurements are the most directly
compatible with the ones presented here based on 
full-disk integrated light X-ray spectra of stars.  

Similarly ``high'' Ne/O ratios have also been seen in $\gamma$-ray
observations of flares, $^3$He-rich solar energetic particle events,
and in the decay phase of long duration soft X-ray
events.\cite{Murphy.etal:91,Ramaty.etal:95,Schmelz:93} The
$\gamma$-ray measurements probe material that is irradiated by
downward-flowing accelerated
particles.\cite{Murphy.etal:91,Ramaty.etal:95} The particle beams
penetrate through to the chromosphere which is likely more
representative of the underlying photospheric material than coronal
regions.

Recent sophisticated models of heliospheric pickup ion and anomalous
cosmic ray populations from the local interstellar medium are also
inconsistent with the current solar Ne/O abundance ratio, but could be
reconciled by raising this to $A_{Ne}/A_O\sim 0.53$,\cite{Frisch.Slavin:03}
which is similar to our values for nearby stars.  This Ne/O ratio is
somewhat higher than the mean from studies of ionized nebulae in the
Milky Way and other galaxies, but falls within the scatter of results
at solar metallicity.\cite{Henry.Worthey:99}

The implication of our study, then, is that the higher of the observed
solar Ne/O abundance ratios are the ones representative of the
underlying solar composition.  This scenario is in accordance
with our observations of Ne/O in nearby stars and reconciles
solar models with helioseismology.

\section*{Methods}

We use the resonance lines of H-like O and of H-like and He-like Ne
to estimate the Ne/O abundance ratio.  
In hot ($10^6$-$10^7$~K) coronal plasma these lines
are formed predominantly by radiative de-excitation of levels excited
by collisions with thermal electrons.  The flux, $F_{ji}$, from such a
transition $j\rightarrow i$ in an ion of an element with abundance $A$
can be written as
\begin{equation}
F_{ji} = A\int_{\Delta T_{ji}} G_{ji}(T) 
\Phi(T) \;dT
\,\,\, \mbox{erg cm$^{-2}$ s$^{-1}$}
\label{e:flux}
\end{equation}
where $G_{ji}(T)$ describes the line {\em emissivity}---the
product of the relative population of the ion in question and the 
excitation rate of the transition as a function temperature, $T$.  The kernel
$\Phi(T)$---the emission measure distribution---describes
the excitation power of the plasma as a 
function of temperature, which is proportional to the mean of the
square of the electron density, $N_e$, and the emitting volume $V$,
$\Phi(T)=\overline{N_e^2}(T)\frac{dV(T)}{dT}$.  

\begin{figure}
\center\includegraphics[width=0.80\textwidth]{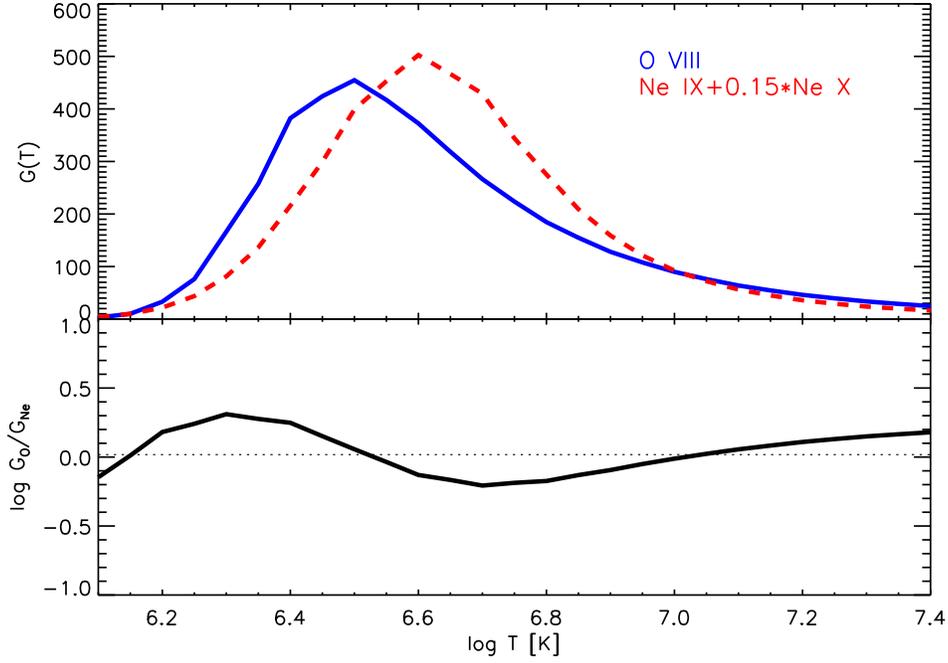}
\caption{The temperature-insensitive O/Ne emissivity ratio.  The upper
panel illustrates the emissivities $G_O$, of the O~VIII $2p\,
^2P_{3/2,1/2} \rightarrow 1s\, ^2S_{1/2}$ line, and $G_{Ne}$ of the
Ne~IX $1s2p\, ^1P_1 \rightarrow 1s^2\, ^1S_0$ and Ne~X $2p\,
^2P_{3/2,1/2} \rightarrow 1s\, ^2S_{1/2}$ lines combined as
$G_{Ne}=G_{NeIX}+0.15G_{NeX}$.  The lower panel shows the logarithmic
ratio $G_O/G_{Ne}$.  Emissivities are based on electron excitation
rates and ion
populations\cite{Mazzotta.etal:98} compiled in the CHIANTI
database,\cite{Young.etal:03} as implemented in
PINTofALE.\cite{Kashyap.Drake:00} }
\label{f:emissrat}
\end{figure}

If we can choose O and Ne lines whose $G_{ji}(T)$ functions have very
similar temperature dependence, an abundance ratio by number,
$A_{Ne}/A_{O}$, can be
derived simply from the ratio of their observed line fluxes, $F_O$ and
$F_{Ne}$, since all the temperature-dependent terms in
Equation~\ref{e:flux} cancel:
\begin{equation}
\frac{A_{Ne}}{A_{O}}=
\overline{\left(\frac{G_O}{G_{Ne}}\right)}\frac{F_{Ne}}{F_{O}}
\end{equation}
An early study of Ne/O ratios in solar active
regions\cite{Acton.etal:75} used the ratio of Ne~IX $1s2p\, ^1P_1
\rightarrow 1s^2\, ^1S_0$ to O~VIII $2p\, ^2P_{3/2,1/2} \rightarrow
1s\, ^2S_{1/2}$.  This ratio does, however, have some significant
residual dependence on temperature.\cite{Acton.etal:75} Here we remove
much of this temperature dependence by addition of Ne~X $2p\,
^2P_{3/2,1/2} \rightarrow 1s\, ^2S_{1/2}$; our combined Ne $G_{ji}(T)$
function is $G_{Ne}=G_{NeIX}+0.15G_{NeX}$. The resulting ratio
$G_O/G_{Ne}$ is illustrated as a function of temperature in
Figure~\ref{f:emissrat}.  We have verified the small residual
temperature sensitivity evident in the lower panel of
Figure~\ref{f:emissrat} to be negligible for our analysis by
integrating the products of $G_O$ and $G_{Ne}$ with
empirically-derived emission measure distributions, $\Phi(T)$, for different
stars,\cite{Drake.etal:01,Garcia-Alvarez.etal:04} and for
functions $\Phi\propto T^a$, with $1 \leq a\leq 4$: the integrated
emissivity ratio from these tests was $\int \Phi G_{O}\, dT/\int \Phi
G_{Ne}\, dT=1.2\pm 0.1$.  We conclude that the line ratio method is
robust and the higher Ne/O abundance ratio found here will not be
significantly changed through performing full emission measure
distribution modelling.

We measured Ne and O line fluxes (listed in Table~1) from {\it
Chandra} HETG X-ray spectra obtained directly from the Chandra public
data archive (http://cda.harvard.edu).  Final listed fluxes for Ne~X
include small reductions ($\leq 15$\% for 17 out of 21 or our stars,
and 25-37\% for the remainder) to account for a weak blend of Fe XVII
at 12.12~\AA .  The Fe~XVII 12.12~\AA\ contribution was estimated by
scaling the observed strengths of unblended Fe~XVII lines at 15.26,
16.77, 17.05 and 17.09~\AA\ (the strong 15.01~\AA\ resonance line was
omitted to avoid potential problems with its depletion through
resonance scattering) by their theoretical line strengths relative to
the 12.12~\AA\ line as predicted by the CHIANTI database.  Minor
blending in the wings of the Ne~IX~13.447~\AA\ line was accounted for
by fitting simultaneously with the neighbouring weaker lines,
comprised of a Fe~XIX-XXI blend at 13.424~\AA\ and Fe~XIX 13.465~\AA,
following a detailed study of these features in the Capella binary
system.\cite{Ness.etal:03} Since these blend corrections are generally
very small, the uncertainties in these procedures have negligible ($<
10$\% ) influence on the derived Ne/O abundance ratios.




\def \apj{{\it Astrophys.\ J.}}
\def \apjl{{\it Astrophys.\ J.\ Lett.}}
\def \apjs{{\it Astrophys.\ J.\ Suppl.\ Ser.}}
\def \prc{{\it Phys.\ Rev.\ C}}
\def \aap{{\it Astron.\ Astrophys.}}
\def \aaps{{\it Astron.\ Astrophys.\ Suppl.}}
\def \mnras{{\it Mon.\ Not.\ R.A.S.}}
\def \physscr{{\it Phys.\ Scripta}}
\def \pasp{{\it Publ.\ Astron.\ Soc.\ Pac.}}
\def \gca{{\it Geochim. Cosmochim.\ Act.}}
\def \aapr{{\it Astron.\ Astrophys.\ Rev.}}

\bibliographystyle{naturemag}


\begin{addendum}
\item[Supplementary Information] Supplementary Information 
accompanies this paper on
www.nature.com/nature 
\item JJD was supported by a NASA contract to the {\em Chandra X-ray
Center}.  PT was supported by a Chandra award issued by Chandra X-ray
Center, which is operated by SAO for and on behalf of NASA.  JJD
thanks the NASA AISRP for providing financial assistance for the
development of the PINTofALE package.  We thank Drs. G.~Share,
R.~Murphy, W.~Ball and D.Garcia-Alvarez for useful discussions and
comments.
\item[Competing interests statement] The authors declare that they have no
competing financial interests.
\item[Correspondence] Correspondence and requests for materials
should be addressed to J.J.~Drake~(email: jdrake@cfa.harvard.edu).
\end{addendum}


\newpage

\begin{table}
\label{t:data}
\caption{Spectral line fluxes and derived Ne/O abundance ratios for
the stars analysed in this study.  Line fluxes were measured from the
Medium Energy Grating (MEG) component of {\it Chandra} HETG spectra by
line profile fitting using the Package for Interactive Analysis of
Line Emission (PINTofALE) software\cite{Kashyap.Drake:00} (freely
available from {\tt http:hea-www.harvard.edu/PINTofALE}).  The
effective collecting area of the instrument was accounted for using
standard {\it Chandra} calibration products and techniques (see
http://cxc.harvard.edu/ciao/ for details).  Ne/O abundance ratios were
obtained assuming the O/Ne line emissivity ratio of $\int \Phi G_{O}\,
dT/\int \Phi G_{Ne}\, dT=1.2\pm 0.1$, as described in Methods. Stated
flux and abundance ratio uncertainties correspond to $1\sigma$ limits.} 
\footnotesize
\begin{tabular}{lccccccl}
\hline
     &   & Dist.& log & \multicolumn{3}{c}{Line Fluxes
     ($10^{-13}$~erg~cm$^{-2}$~s$^{-1}$)} & \\ \cline{5-7}
Star & Type & [pc] & ($L_X/L_{bol}$) & $F_{NeIX}$ & $F_{NeX}$ & $F_{OVIII}$ &
     $A_{Ne}/A_O$ \\ \hline
AU~Mic		&  M1V           &  9.9  &  -3.37  &  3.08 $\pm$ 0.20  &   5.60 $\pm$ 0.24  &  10.4  $\pm$ 0.4   &   0.38 $\pm$ 0.05 \\
Prox~Cen	&  M5Ve         &  1.3  &  -3.92  &  0.70 $\pm$ 0.14  &   1.22 $\pm$ 0.4   &   3.2  $\pm$ 0.3   &   0.28 $\pm$ 0.22 \\
EV~Lac		&  M1.5V        &  5.1  &  -3.11  &  3.09 $\pm$ 0.10  &   3.99 $\pm$ 1.0   &   9.16 $\pm$ 0.20  &   0.40 $\pm$ 0.19 \\
AB~Dor		&  K0V          &  15   &  -3.26  &  4.9  $\pm$ 0.4   &   10.7 $\pm$ 0.4   &  14.6  $\pm$ 0.5   &   0.44 $\pm$ 0.06 \\
HR9024		&  G1III        & 135   &  -3.65  &  0.30 $\pm$ 0.11  &   2.09 $\pm$ 0.18  &   2.40 $\pm$ 0.23  &   0.26 $\pm$ 0.15 \\
31~Com		&  G0III        &  94   &  -4.70  &  0.21 $\pm$ 0.05  &   0.84 $\pm$ 0.08  &   1.30 $\pm$ 0.17  &   0.27 $\pm$ 0.13 \\
$\beta$~Cet	&  K0III        &  29.4 &  -5.35  &  2.15 $\pm$ 0.09  &   5.30 $\pm$ 0.14  &   8.0  $\pm$ 0.3   &   0.37 $\pm$ 0.04 \\
Canopus		&  F0II         &  96   &  -7.23  &  0.52 $\pm$ 0.10  &   0.76 $\pm$ 0.05  &   1.33 $\pm$ 0.25  &   0.48 $\pm$ 0.15 \\
$\mu$~Vel	&  G5III/...    &  35.5 &  -5.46  &  1.40 $\pm$ 0.13  &   1.80 $\pm$ 0.24  &   4.2  $\pm$ 0.3   &   0.40 $\pm$ 0.11 \\
Algol		&  B8V/K1IV     &  28   &  -3.5   &  2.9  $\pm$ 0.3   &  12.8  $\pm$ 0.7   &  13.8  $\pm$ 0.5   &   0.35 $\pm$ 0.09 \\
ER~Vul		&  G0V/G5V      &  50   &  -3.45  &  1.10 $\pm$ 0.11  &   3.0  $\pm$ 0.19  &   3.8  $\pm$ 0.3   &   0.41 $\pm$ 0.10 \\
44~Boo		&  G1V/G2V      &  13   &  -4.16  &  3.75 $\pm$ 0.29  &   8.1  $\pm$ 0.4   &  15.2  $\pm$ 0.6   &   0.33 $\pm$ 0.06 \\
TZ~CrB		&  G0V/G0V      &  22   &  -3.36  &  7.63 $\pm$ 0.30  &  15.5  $\pm$ 0.4   &  22.1  $\pm$ 0.5   &   0.45 $\pm$ 0.04 \\
UX~Ari		&  G5V/K0IV     &  50   &  -3.63  &  3.62 $\pm$ 0.27  &  12.4  $\pm$ 0.3   &   9.7  $\pm$ 0.5   &   0.56 $\pm$ 0.08 \\
$\xi$~UMa	&  G0V/G0V      &  7.7  &  -4.26  &  4.56 $\pm$ 0.24  &   5.41 $\pm$ 0.22  &  15.8  $\pm$ 0.6   &   0.34 $\pm$ 0.03 \\
II~Peg		&  K2V/...      &  42   &  -2.92  &  5.26 $\pm$ 0.30  &  19.4  $\pm$ 0.5   &  19.4  $\pm$ 0.8   &   0.42 $\pm$ 0.06 \\
$\lambda$~And	&  G8III/...    &  26   &  -4.53  &  2.79 $\pm$ 0.20  &   8.82 $\pm$ 0.23  &  10.3  $\pm$ 0.4   &   0.40 $\pm$ 0.05 \\
TY~Pyx		&  G5IV/G5IV    &  56   &  -3.55  &  2.09 $\pm$ 0.22  &   4.78 $\pm$ 0.09  &   4.9  $\pm$ 0.5   &   0.57 $\pm$ 0.08 \\
AR~Lac		&  G2IV/K0IV    &  42   &  -3.72  &  2.7  $\pm$ 0.3   &   9.70 $\pm$ 0.28  &   9.3  $\pm$ 0.5   &   0.45 $\pm$ 0.07 \\
HR1099		&  G5IV/K1IV    &  29   &  -3.29  &  8.36 $\pm$ 0.26  &  27.6  $\pm$ 0.3   &  27.9  $\pm$ 0.5   &   0.45 $\pm$ 0.03 \\
IM~Peg		&  K2III-II/... &  97   &  -3.98  &  3.00 $\pm$ 0.29  &  15.8  $\pm$ 0.3   &  13.0  $\pm$ 0.7   &   0.41 $\pm$ 0.06 \\
\hline
\end{tabular}
\end{table}

\newpage

\end{document}